\documentclass[aps,prl,twocolumn,superscriptaddress]{revtex4}
\usepackage{graphicx,amsmath,amssymb}

\begin{document}

\title{Resonant excitation of graphene K-phonon and intra-Landau level excitons in magneto-optical spectroscopy.}


\author{M. Orlita$^{*}$}
\affiliation{Laboratoire National des Champs Magn$\acute{e}$tiques
Intenses, CNRS-UJF-UPS-INSA, B.P. 166, 38042 Grenoble Cedex 9,
France} \affiliation  {Institute of Physics, Charles University,
Ke Karlovu 5, CZ-121 16 Praha 2, Czech Republic}

\author{Liang Z. Tan}
\affiliation{Department of Physics, University of California at
Berkeley, and Materials Sciences Division, Lawrence Berkeley National
Laboratory, Berkeley, CA 94720}

\author{M. Potemski}
\affiliation{Laboratoire National des Champs Magn$\acute{e}$tiques
Intenses, CNRS-UJF-UPS-INSA, B.P. 166, 38042 Grenoble Cedex 9,
France}

\author{M. Sprinkle}
\affiliation{School of Physics, Georgia Institute of Technology,
Atlanta, Georgia 30332, USA}

\author{C. Berger}
\affiliation{School of Physics, Georgia Institute of Technology,
Atlanta, Georgia 30332, USA} \affiliation{Institut N$\acute{e}$el,
CNRS-UJF B.P. 166, 38042 Grenoble Cedex 9, France}

\author{W. A. de Heer}
\affiliation{School of Physics, Georgia Institute of Technology,
Atlanta, Georgia 30332, USA}

\author{Steven G. Louie$^{*}$}
\affiliation{Department of Physics, University of California at
Berkeley, and Materials Sciences Division, Lawrence Berkeley National
Laboratory, Berkeley, CA 94720}

\author{G. Martinez}
\affiliation{Laboratoire National des Champs Magn$\acute{e}$tiques
Intenses, CNRS-UJF-UPS-INSA, B.P. 166, 38042 Grenoble Cedex 9,
France}

\date{\today}
\begin{abstract}
Precise infra-red magneto-transmission experiments have been
performed in magnetic fields up to 32~T on a series of multi-layer
epitaxial graphene samples. We observe changes in the spectral
features and broadening of the main cyclotron transition when the
incoming photon energy is in resonance with the lowest Landau
level separation and the energy of a  $K$ point optical phonon. We
have developed a theory that explains and quantitatively
reproduces the frequency and magnetic field dependence of the
phenomenon as the absorption of a photon together with the
simultaneous creation of an intervalley, intra-Landau level
exciton and a $K$-phonon.

\end{abstract}

\maketitle

Phonons, electrons, holes and plasmons are fundamental elementary
excitations in solids, and their interactions among themselves and
with light strongly influence the electronic and optical
properties of a system. Graphene, a single atomic layer of carbon,
exhibits many fascinating properties of basic and practical
interest. Its low-energy charge carriers behave like
two-dimensional (2D) massless Dirac fermions. The unique
electronic structure of graphene has led to the prediction and
observation of a number of novel phenomena involving various
multi-elementary excitation interactions including
electron-photon, electron-phonon, electron-electron, and
electron-plasmon couplings. Recent discoveries include the
observation of plasmarons in ARPES \cite{plasmarons}, controlling
of quantum pathways in inelastic light scattering \cite{Feng1},
and strong excitonic effects in optical absorption
\cite{Yang,Geim1,Heinz}, among others.

In this paper, we report the discovery of another intriguing
phenomenon seen in the magneto-optical response of graphene when
the incoming photon energy is in resonance with the lowest Landau
level separation and the energy of a Brillouin-zone-boundary
phonon at the $K$ point. The frequency and magnetic field
dependence of the spectral features are understood and explained
in terms of resonant transitions which involve the absorption of a
photon together with the simultaneous creation of an intervalley,
intra-Landau level exciton and a $K$-phonon. The phenomenon
provides a novel manifestation of multi-elementary excitation
interactions in condensed matter.

The phonons that interact the most strongly with electrons in
graphene are those at the Brillouin-zone-center ($\Gamma$) and at
the Brillouin-zone-boundary ($K$) \cite{AndoKPhonon}. The
$\Gamma$-phonon has been the subject of theoretical study
\cite{AndoKPhonon} and its signatures have been observed in
several experiments\cite{Ferrari,Bostwick}. The $K$-phonon also
interacts strongly with electrons, giving rise to characteristic
features in Raman spectra \cite{Ferrari}. Compared to
$\Gamma$-phonons, the effect of $K$-phonons on optical experiments
is more subtle and indirect, due to the large difference in
momenta between photons and the $K$-phonons. The large wavevector
of the $K$-phonons may also couple the two inequivalent valleys in
the graphene band structure, introducing an additional degree of
freedom into the system. The present study shows that this valley
degree of freedom in the coupling has unique and important
consequences in the magneto-optical response of graphene.

In our experiment, precise infra-red transmission measurements were
performed on multi-layer epitaxial graphene samples, at 1.8 K, under
magnetic fields up to 32~T.
It is known that under a perpendicular magnetic field $B$, the electronic
structure of graphene is quantized to discrete Landau levels (LL) with the
energy $E_{n}=sgn(n)v_{F}\sqrt{2e\hbar B|n|}$ where $n$ are integers
including 0. Multiple optical
transitions are allowed and have been observed \cite{Sadowski}
between LL $|n|$ to $|n|\pm1$ if these levels have
appropriate occupation factors. We focus our attention on the
magnetic field evolution of the transmission spectrum near the main
optical absorption peak, with
energy $E_{01}=v_{F}\sqrt{2e\hbar B}$ involving the $n=0$ LL (i.e.,
transitions from $n=-1$ to $n=0$ or from $n=0$ to $n=1$.) It is
discovered that a strong change in the optical spectrum occurs when
$E_{01}$ reaches an energy near
the energy of a $K$-phonon. This is interpreted as a signature of a
new multi-elementary excitation phenomenon involving interaction
between electrons and the $K$-phonon, and a theory is
developed to explain quantitatively the phenomenon.

\begin{figure*}
\includegraphics[width=2.0\columnwidth]{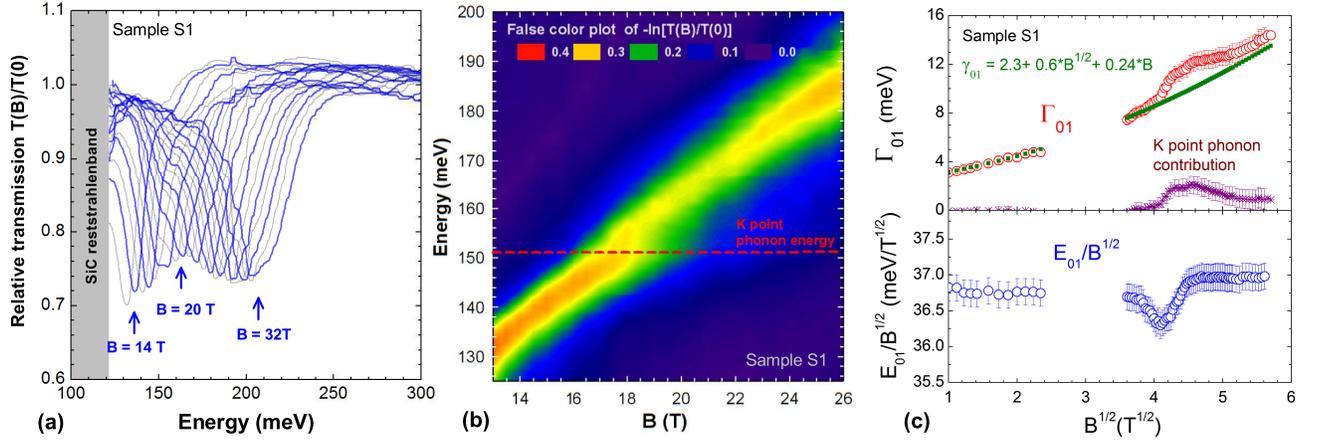}
\caption{\label{Fig1}  (a)  Relative transmission spectra
(T(B)/T(0)) of sample S1, for different magnetic field values. (b)
Same data as in (a) presented in a false color plot of
-ln[T(B)/T(0)]. (c) Bottom panel: variation of cyclotron
transition energy $E_{01}/B^{1/2}$ as a function of $B^{1/2}$
(open blue circles). Top panel: variation of the fitted linewidth
$\Gamma_{10}(B)$ of the transition $E_{01}(B)$ as a function of
$B^{1/2}$ (open red circles).This variation is de-convoluted is
two parts, one named $\gamma_{01}(B)$ (green dots) and the
remaining part (stars) which is assigned to the $K$-phonon
contribution.}
\end{figure*}

A series of  four multi-layer epitaxial graphene samples were
grown \cite{Berger} on the C-terminated surface of SiC. The
thickness of the SiC substrate has been reduced to ~$60 \mu m$ in
order to minimize the very strong double-phonon absorption of SiC
in the energy range of interest. Whereas the transmission spectra
reveal some contributions of graphite Bernal-stacked inclusions,
the major part displays the electronic band structure of an
isolated graphene monolayer resulting from the characteristic
rotational stacking of the sheets \cite{Hass}. Using standard
techniques  \cite{Faugeras}, we measure the relative transmission
spectra TR($B,\omega$)(see Supplementary Information SI-I).

The TR spectra of the sample S1 are displayed in Fig.~\ref{Fig1}a
and b for different values of $B$. They show, as for all samples,
an apparent splitting of the transmission dip into two and a
pronounced increase of the broadening of the dip, as well as a
change in the $B$-field dependence of the overall position
$E_{01}$ for fields larger than 17~T.

In general, the detailed analysis of such spectra requires the use
of a multi-layer dielectric model including all layer dielectric
properties of the sample. In particular, for each graphene sheet,
one has to introduce the corresponding components of the optical
conductivity tensor $\sigma_{xx}(\omega)$ and
$\sigma_{xy}(\omega)$, defined by
$J_{\alpha}=\sum_\beta\sigma_{\alpha\beta}\mathcal{E}_{\beta}$
($J_{\alpha}$ and $\mathcal{E}_{\beta}$ denote current density and
electric field vectors respectively). Here, the $x$ and $y$-axis
lie in the plane of the sample. For instance
$\sigma_{xx}(\omega)$, for transitions involving the n=0 LL, is
written as:

\begin{equation}\label{sigmaxx}
\sigma_{xx}(\omega,B)= \imath
\frac{e^{3}B}{h\omega}\sum_{r,s}\frac{M^2_{r,s}(f_{r}(B)-f_{s}(B))}{\hbar\omega-E_{r,s}(B)+\imath\Gamma_{rs}(B)}
\end{equation}

\noindent where $r$,$s$ scan the values 0 and $\pm 1$, $0\leq
f_{r}\leq 1$ is the occupation factor of the LL $r$, $M_{r,s}$ the
optical matrix element, $E_{r,s}=E_{r}-E_{s}=E_{01}$ and
$\Gamma_{rs}(B)=\Gamma_{01}(B)$ measures the broadening of the
transition .
However this approach requires the
knowledge of the number of effective active layers as well as
their carrier densities  which, in turn, implies some
approximations (see SI-II for details). We will use this approach
\textit{in a refined analysis}, but in our initial analysis, fit directly the
$E_{01}$ transition with a single Lorentzian line.

This fit provides, for each value of $B$, two independent
parameters $E_{01}(B)$ and $\Gamma_{01}(B)$ which are plotted in
Fig.~\ref{Fig1}c as a function of $B^{1/2}$. The variation of
$\Gamma_{01}(B)$ shows clearly an extra bump beyond 17~T (clearly
apparent in Fig.~\ref{Fig1}b )  whereas that of
$E_{01}(B)/B^{1/2}$ exhibits a downwards kink at the same field.
The evolution of $\Gamma_{01}(B)$ is decomposed in two parts: one
named $\gamma_{01}(\textrm{meV})= 2.3+ 0.6\sqrt{B}+0.24 B$ (full
dots) and an extra contribution to $\Gamma_{01}$ represented by
stars in the top panel of Fig.~\ref{Fig1}c. The same decomposition
is obtained for all samples with slightly different coefficients
which are discussed in SI-II. The $\sqrt{B}$ dependence of
$\gamma_{01}$ is attributed to scattering with short range
impurities \cite{impurities}; the linear variation with $B$ is not
predicted by scattering mechanisms considered in previous studies
\cite{impurities} and will be discussed later. The shape of the
extra contribution to $\Gamma_{01}$ has a threshold at around 17~T
($\sim 151 ~\textrm{meV}$) with a maximum amplitude of about $2
~\textrm{meV}$ for all samples, within the experimental
uncertainties, before decreasing at higher fields. In that energy
range the only likely excitations in graphene which could play a
role are the zone-boundary $K$-phonons \cite{Maultzsch,Gruneis}.
Note that, at this threshold energy, the
 evolution $E_{01}/B^{1/2}$  (bottom panel of
Fig.~\ref{Fig1}c) is the signature of interactions because, in the
absence of such interactions, it should be an horizontal line.
This is reminiscent of the response function of the Fr\"{o}hlich
interaction observed in polar semiconductors
\cite{Faugeras,Orlita} where a zone-center longitudinal optical
phonon is emitted when the energy of the cyclotron transition
exceeds that of the phonon. However, graphene is not a polar
material and, furthermore, we are not expecting a direct
interaction of the $K$-phonon with the infra-red light. One
therefore has to explain why and how the $K$-phonons enter the
interaction.
\begin{figure}
\includegraphics*[width=1.0\columnwidth]{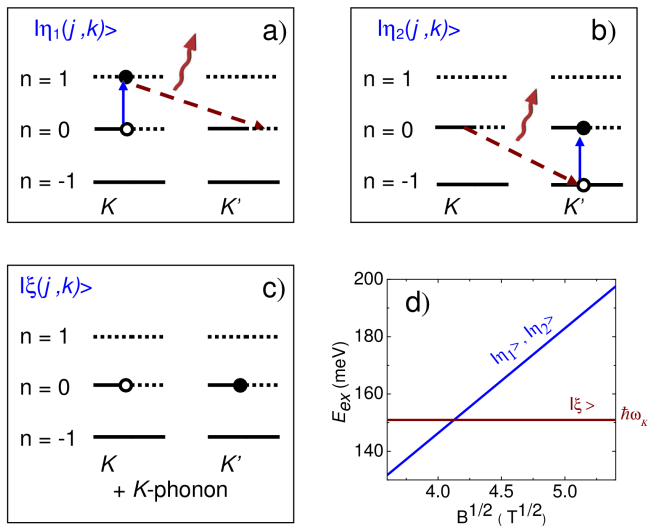}
\caption{\label{Fig2} Cyclotron resonant excitations
$\lvert\eta_1(j,k)\rangle$(a) and $\lvert\eta_2(j,k)\rangle$ (b)
involving the $n=0$ Landau levels, at either the $K$ or $K'$
valley, with the excitation energy $E_{ex}=E_{01}$. (c) Excited
state $\lvert\xi_{\vec{q}}(j,k)\rangle$ corresponding to
intervalley transfer of an electron within the $n=0$ LL from
$\vec{K}$ to $\vec{K}'$ and creation of a phonon with wavevector
$\vec{K}+\vec{q}$, which has excitation energy
$E_{ex}=\hbar\omega_{\vec{K}+\vec{q}}$ . (d) At resonance, i.e.
$E_{01}=\hbar\omega_{\vec{K}+\vec{q}}$, $\lvert\eta_1\rangle$ and
$\lvert\eta_2\rangle$ may be transformed to $\lvert\xi\rangle$ via
an emission of a $\vec{K}+\vec{q}$ phonon as indicated by the
dashed brown lines. The mixing between the three excited states,
mediated by the electron-phonon interaction,  is the strongest
when the resonance condition is satisfied
($E_{01}=\hbar\omega_{\vec{K}+\vec{q}}$) . Electron-phonon
scatterings involving LL $\lvert{}n\rvert\ge2$ do not result in
similar resonances in the experimental energy range (see SI-IV.)
In general, there are scattering processes involving phonons of
different $\vec{q}$ vectors near $K$, but we find the coupling
strength  significant only in a small region with
$\vec{q}\approx0$ where the phonon energy is essentially constant
and equal to $\hbar\omega_K$. The theoretical description of this
continuum of near $K$ phonons may be incorporated effectively by
considering a single state $\lvert\xi\rangle$ with a complex self
energy and using an appropriately renormalized coupling constant
between $\lvert\eta_1\rangle$, $\lvert\eta_2\rangle$ and
$\lvert\xi\rangle$ (see derivation in SI-IV) .}
\end{figure}

The mechanism for the interaction of the cyclotron transition with
the $K$-phonon is not \emph{a priori} clear for the reason of
conservation of wavevector, i.e., a state corresponding to an
optical cyclotron transition which has wavevector $0$ cannot decay
into a state with a phonon of wavevector $\overrightarrow{K}$
without simultaneously emitting another entity of wavevector
$-\overrightarrow{K}$. Given that this interaction must
necessarily involve the electron-phonon interaction, and that the
low-energy electronic states at the two valleys of graphene are
separated in reciprocal space by a wavevector $\overrightarrow{K}$,
we propose that an intervalley intra-LL electron scattering
process (creation of a zero-energy intervalley electron-hole pair)
is responsible for the observed phenomenon (Fig.~\ref{Fig2}.)

To explain the observed spectral features, we shall consider the
coupling of the photo-excited (cyclotron transition) states
(denoted as $\lvert\eta_1(j,k)\rangle$ and $\lvert\eta_2(j,k)\rangle$
for the $0\rightarrow1$ and $-1\rightarrow0$ LL transitions
respectively) with an excited state of the system containing an
intervalley, intra-Landau level electron-hole pair excitation and a
$K$-phonon of wavevector $\vec{K}+\vec{q}$ (denoted as
$\lvert\xi_{\vec{q}}(j,k)\rangle$). See Fig.\ 2. Here $j$ is the valley index and $k$
is the quantum number which describes the degenerate states within a LL.
As illustrated in Fig.\ 2, at cyclotron transition energy $E_{01}$ near
$\hbar\omega_{\vec{K}+\vec{q}}$, $\lvert\eta_1(j,k)\rangle$ and
$\lvert\eta_2(j,k)\rangle$ are connected to
$\lvert\xi_{\vec{q}}(j,k)\rangle$ via an emission of $\vec{K}+\vec{q}$
phonons. The electron-phonon coupling in the system can then strongly
mix these states at resonance and lead to significant changes in the
absorption spectrum.

In order to reproduce quantitatively the transmission spectra, one
has now to use the multi-layer dielectric model (see SI-II for
details). The optical conductivity is calculated using the Green's
function formalism introduced by Toyozawa \cite{Toyozawa1}. The
conductivity components are
\begin{equation}\label{green}
\begin{aligned}
\textrm{Re}\,\sigma_{ij}(\hbar\omega)&\propto\textrm{Im}\langle{}0\lvert{}M_{i}G(\hbar\omega)M_{j}\rvert{}0\rangle
\end{aligned}
\end{equation}
\noindent where $\lvert{}0\rangle$ is the ground state of the
system, and $M_{i}$, $M_{j}$ are row and column vectors containing
optical matrix elements. $G$ is the retarded Green's function
written here as
$G(\hbar\omega)=[\hbar\omega-H-\Sigma(\omega)+is]^{-1}$ with
$s\rightarrow 0^-$ with $H$ being a 3x3 Hamiltonian describing the
interaction between the three states $\lvert\eta_1\rangle$,
$\lvert\eta_2\rangle$ and $\lvert\xi\rangle$. The poles of $G$ are
 broadened by a self energy $\Sigma(\omega)$ reflecting
additional lifetime effects due to the $K$-phonon dispersion and
the excitations from electron-acoustic phonons scattering,
electron-electron interactions and other environmental effects not
captured by $H$ (see SI-V). In the theoretical calculations, we have
incorporated this additional physics near resonance with a broadening function of the form
$\Gamma_{01}(\omega,B)=\gamma_{01}(B)+\textrm{Im}(\Sigma(\omega))$
with $\gamma_{01}(B)$ given above and $\textrm{Im}\,(\Sigma)$
taken to be
$\textrm{Im}(\Sigma(\omega))=\pi{}A^{2}f(R/\hbar\omega_{K})\theta(\hbar\omega-\hbar\omega_K)e^{-R(\omega/\omega_K-1)}$.
The derivation of the form of $\textrm{Im}\,(\Sigma(\omega))$ within our
theoretical model is given in SI-IV and SI-V.
$A^{2}(~\textrm{meV}^2)=1.94 \times B(\textrm{T})$ is the
electron-phonon parameter determined from Ref. \cite{Gruneis}. $f$
is a factor, dependent on the relative magnitudes of the valley
($\Delta_V$) and spin ($\Delta_S$) spittings of the $n=0$ LL, that
is determined by the occupations of the spin and valley sublevels
of the $n=0$ LL (see SI-IV). The threshold structure of
$\textrm{Im}(\Sigma(\omega))$ represents the onset of $K$-phonon
emission at energies above $\hbar\omega_{K}$. The only parameters
in the theory are the phonon energy $\hbar\omega_K$ and the factor
$R$. The absorption spectra can be obtained from
$\textrm{Re}\,\sigma(\omega)$ by a simple proportionality relation
(see SI-V and \cite{Sadowski}).

\begin{figure}
\includegraphics*[width=0.9\columnwidth]{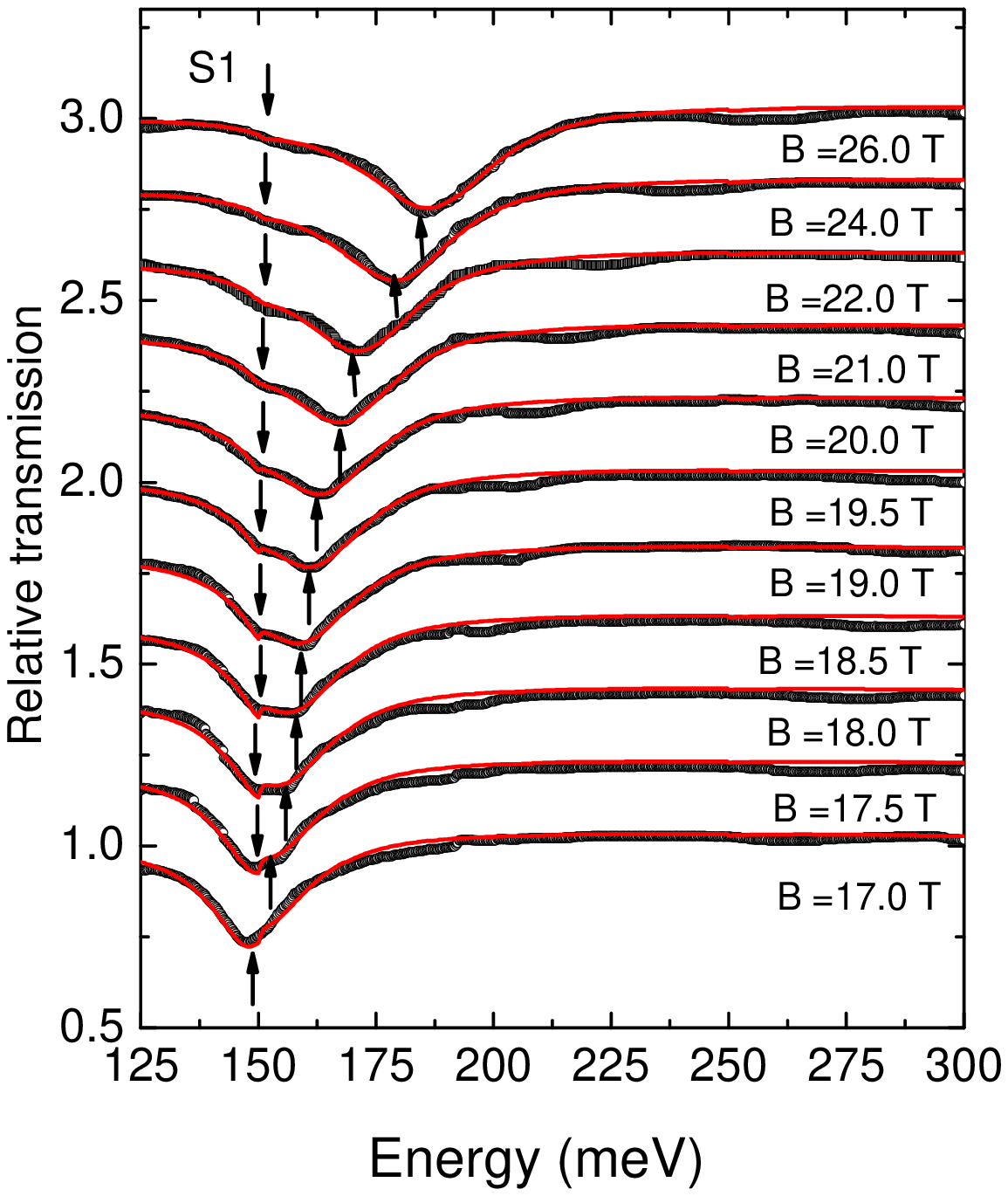}
\caption{\label{Fig3} Comparison, for different values of the
magnetic field, of experimental data for sample S1 (open black
dots) with the theoretical curves (red lines) using the proposed
K-phonon interaction model. As a guide for the eye, upward arrows
show the evolution of the main cyclotron resonance line whereas downward arrows
point the structure corresponding to the emission of $K$-phonons.
}
\end{figure}

\begin{figure}
\includegraphics*[width=1.0\columnwidth]{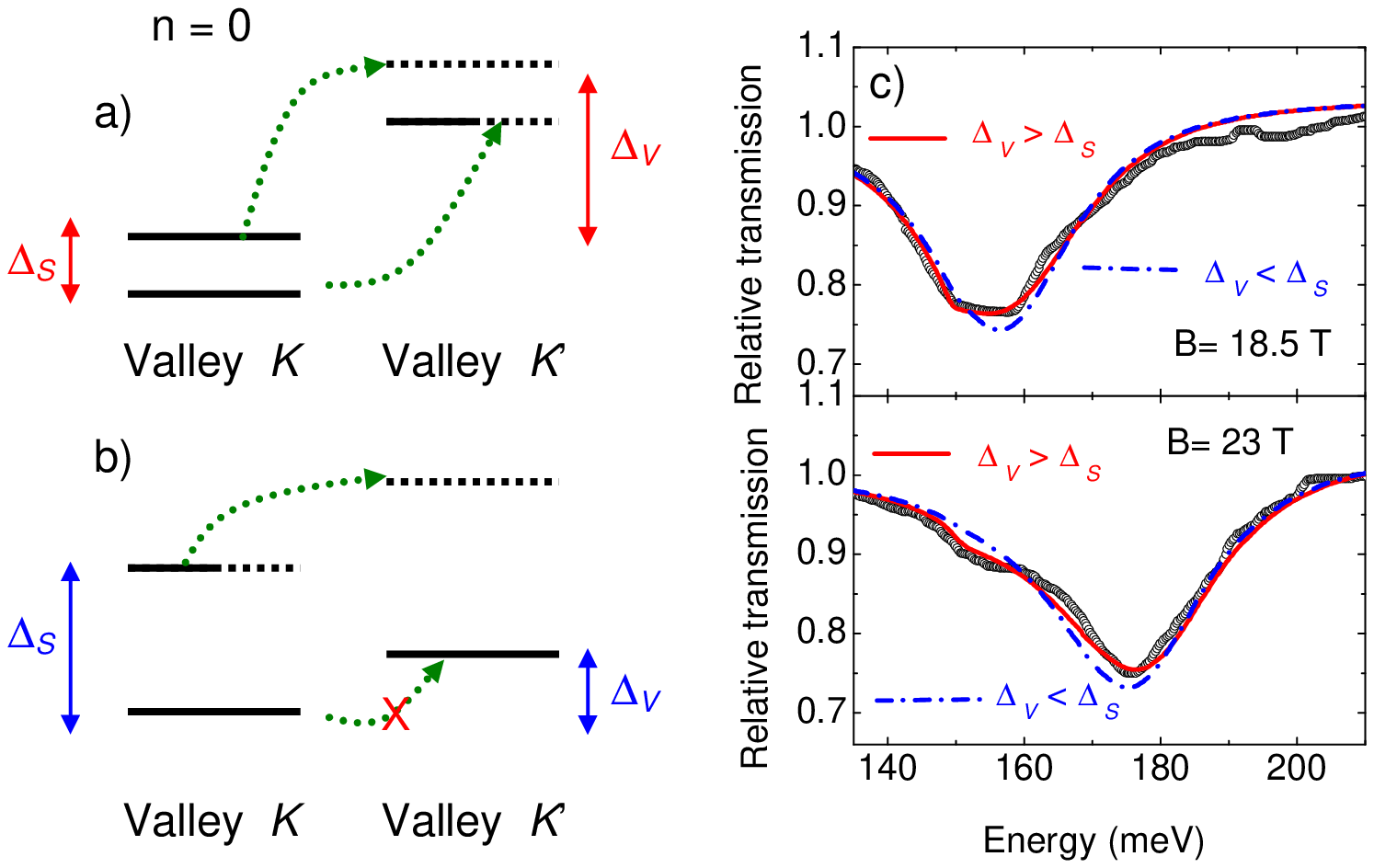}
\caption{\label{Fig4} (a), (b) The spin and valley sublevels of
the $n=0$ LL, for the case when  $\Delta_V>\Delta_S$ and
$\Delta_V<\Delta_S$ respectively. In the former case, intervalley
electronic transitions are possible for both spin orientations. In
the latter case, only one spin orientation supports intervalley
electronic transitions, owing to the Fermi factor. The effective
coupling strength is therefore  different in the  two cases giving
rise to different predictions for the transmission spectra. (c) A
comparison between experimental spectra  and theoretical
simulations for
 $\Delta_V>\Delta_S$ and $\Delta_V<\Delta_S$  for values of magnetic field $B=18.5\textrm{T}$ and
$B=23.0\textrm{T}$ respectively. For all our
samples the splitting of the valley degeneracy is larger than the spin
splitting. This is consistent with recent tunnelling spectroscopy
experiments on graphene on SiC \cite{Song}. Theoretical discussions on
the relative magnitudes of these splittings can be found in Refs.
\cite{spinvalley1,spinvalley2,spinvalley3,Fuchs,spinvalley4,spinvalley5}. }
\end{figure}

In Fig.~\ref{Fig3} we compare the experimental data, for sample
S1, with the calculated results from the model with
$\Delta_V>\Delta_S$. We have taken $\hbar\omega_K=151
~\textrm{meV}$ and $R=3$. The Fermi velocity
$v_{F01}=1.012\times10^6 ~\textrm{ms}^{-1}$ for the transitions
between the $n=0$ LL and $n=\pm1$ LL has been determined
experimentally from the positions of the cyclotron transition
lines (see SI-II and SI-III.) The same parameters have been used
for all the graphene samples (see SI-III).  The good agreement
between theory and experiment, for all samples, lends strong
support to the physical mechanism proposed for the phenomenon. It
is , however only obtained when assuming $\Delta_V>\Delta_S$
(Fig.~\ref{Fig4}).

An interpretation of the linear term in $\gamma_{01}(B)$ is that it may
be a broadening resulting from the breaking of the valley
degeneracy (Fig.~\ref{Fig4}), consistent with the prediction that $\Delta_V\propto
B$, according to several theoretical models
\cite{spinvalley1,spinvalley2,spinvalley3}. In this picture, the
values of $E_{01}$ at the two valleys differ by $\Delta_V$
\cite{Fuchs}, resulting in a broadening of the main line by a
similar amount. An inspection of the linear term in
$\gamma_{01}(B)$ shows that it is greater than the value of
$\Delta_S$ due to Zeeman splitting. This is consistent with our
conclusion that $\Delta_V>\Delta_S$ in our samples (Fig.~\ref{Fig4}). However,
further investigations are needed to test if the valley splitting
is indeed the main contributor to the linear term in
$\gamma_{01}$.

To conclude, we have observed splitting of spectral features and broadening of the main
absorption line above
17~T in our infra-red transmission experiments on epitaxial
graphene that we attribute to a resonance between cyclotron transitions
and K-phonon emission coupled with a zero-energy intervalley
electron-hole excitation. We
have developed a theoretical model that reproduces the experimental results
with quantitative agreement, thereby shedding light on the nature
of the optical transitions and electron-phonon interaction in graphene under strong
magnetic fields. Our results indicate that the valley splitting of the
LL is
larger than the spin splitting in our samples. To our knowledge,
this is the first instance where information on the sublevel
structure of graphene has been obtained using infra-red
transmission measurements, complementing transport- \cite{Kim,
Andrei} and tunnelling-based \cite{Song} experiments.

We thank Cheol-Hwan Park for useful discussions. L.Z.T. and the
theoretical analysis were supported by the Director, Office of
Science, Office of Basic Energy Sciences, Materials Sciences and
Engineering Division, U.S. Department of Energy under Contract No.
DE- AC02-05CH11231. Numerical simulations were supported in part
by NSF Grant DMR10-1006184. Computational resources were provided
by NSF through TeraGrid resources at NICS and by DOE at Lawrence
Berkeley National Laboratory's NERSC facility.This work has been
supported in part by by EuroMagNET II under EU Contract No.
228043. This work has also been supported by projects GACR
P204/10/1020, GRA/10/E006 (Eurographene-EPIGRAT) and IRMA.

\bibliography{basename of .bib file}

\end{document}


\title{Supplementary information for
Resonant excitation of graphene K-phonon and intra-Landau level
excitons in magneto-optical spectroscopy.}


\author{M. Orlita$^{*}$}
\affiliation{Laboratoire National des Champs Magn$\acute{e}$tiques
Intenses, CNRS-UJF-UPS-INSA, B.P. 166, 38042 Grenoble Cedex 9,
France} \affiliation  {Institute of Physics, Charles University,
Ke Karlovu 5, CZ-121 16 Praha 2, Czech Republic}

\author{Liang Z. Tan}
\affiliation{Department of Physics, University of California at
Berkeley, and Materials Sciences Division, Lawrence Berkeley National
Laboratory, Berkeley, CA 94720}

\author{M. Potemski}
\affiliation{Laboratoire National des Champs Magn$\acute{e}$tiques
Intenses, CNRS-UJF-UPS-INSA, B.P. 166, 38042 Grenoble Cedex 9,
France}

\author{M. Sprinkle}
\affiliation{School of Physics, Georgia Institute of Technology,
Atlanta, Georgia 30332, USA}

\author{C. Berger}
\affiliation{School of Physics, Georgia Institute of Technology,
Atlanta, Georgia 30332, USA} \affiliation{Institut N$\acute{e}$el,
CNRS-UJF B.P. 166, 38042 Grenoble Cedex 9, France}

\author{W. A. de Heer}
\affiliation{School of Physics, Georgia Institute of Technology,
Atlanta, Georgia 30332, USA}

\author{Steven G. Louie$^{*}$}
\affiliation{Department of Physics, University of California at
Berkeley, and Materials Sciences Division, Lawrence Berkeley National
Laboratory, Berkeley, CA 94720}

\author{G. Martinez}
\affiliation{Laboratoire National des Champs Magn$\acute{e}$tiques
Intenses, CNRS-UJF-UPS-INSA, B.P. 166, 38042 Grenoble Cedex 9,
France}

\date{\today}

\pacs{78.30.Fs, 71.38.-k, 78.66.Fd}

\maketitle

\section{Experimental details}

Far infra-red magneto-optical experiments have been performed at
magnetic field strengths up to 29-32~T and at a fixed temperature
of 1.8 K. The light (provided and analyzed by a Fourier transform
spectrometer) was delivered to the sample by means of light-pipe
optics. All experiments were performed with nonpolarized light, in
the Faraday geometry with the wave vector of the incoming light
parallel to the magnetic field direction and perpendicular to the
plane of the samples. A Si bolometer was placed directly beneath
the sample to detect the transmitted radiation. The response of
this bolometer is strongly dependent on the magnetic field.
Therefore, in order to measure the absolute transmission TA
($B,\omega)$, we used a sample-rotating holder and measure for
each value of $B$ a reference spectrum through a hole. These
spectra are normalized in turn with respect to TA($0,\omega$) to
obtain a relative transmission spectrum TR($B,\omega$) which only
displays the magnetic field dependent features. Those spectra are
presented in Fig.~\ref{Fig1}.

\section{Global analysis of the spectra}

\begin{figure*}
      \scalebox{0.9}{\includegraphics{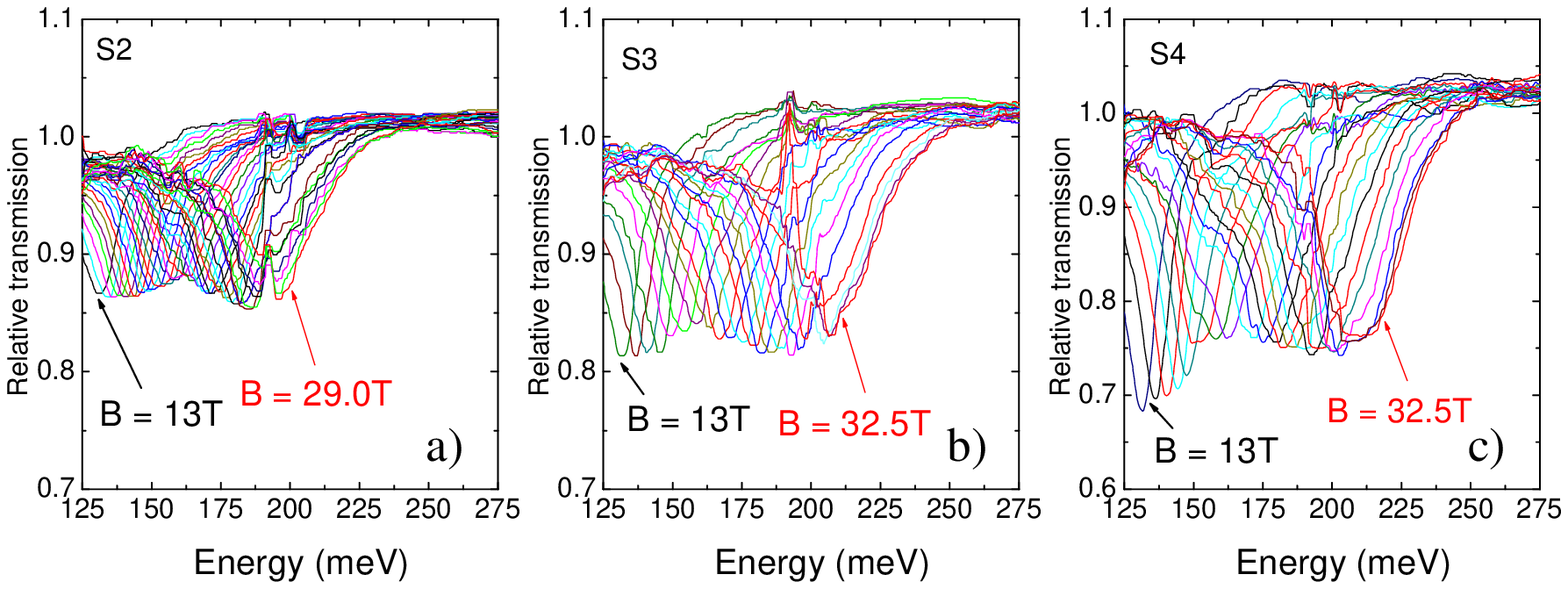}}
          \caption{\label{Fig1}   Relative
transmission spectra, as a function of the energy,  for different
values of the magnetic field $B$ : (a) sample S2, (b) sample S3
 and (c) Sample S4. }
\end{figure*}

In Fig.~\ref{Fig1}, the relative transmission spectra are shown
for sample S2 (a), sample S3 (b) and sample S4 (c). In general,
for filling factors $\nu<2$, the oscillator strength of a
magneto-optical transition in graphene is expected to be
proportional to $\sqrt{B}$ if the broadening parameter
$\Gamma_{01}$ is a constant. If $\Gamma_{01}\sim \sqrt{B}$ the
minimum of the transmission should be independent on $B$.  However
and as for the sample S1 (Fig.~\ref{Fig1}(a) of the main text),
all the observed transmission data  display an extra broadening of
the $E_{01}$ transition for magnetic fields values larger than
17~T. This leads us to conclude that such an effect is an
intrinsic property of the graphene sheets.

\begin{figure*}
\includegraphics*[width=0.9\columnwidth]{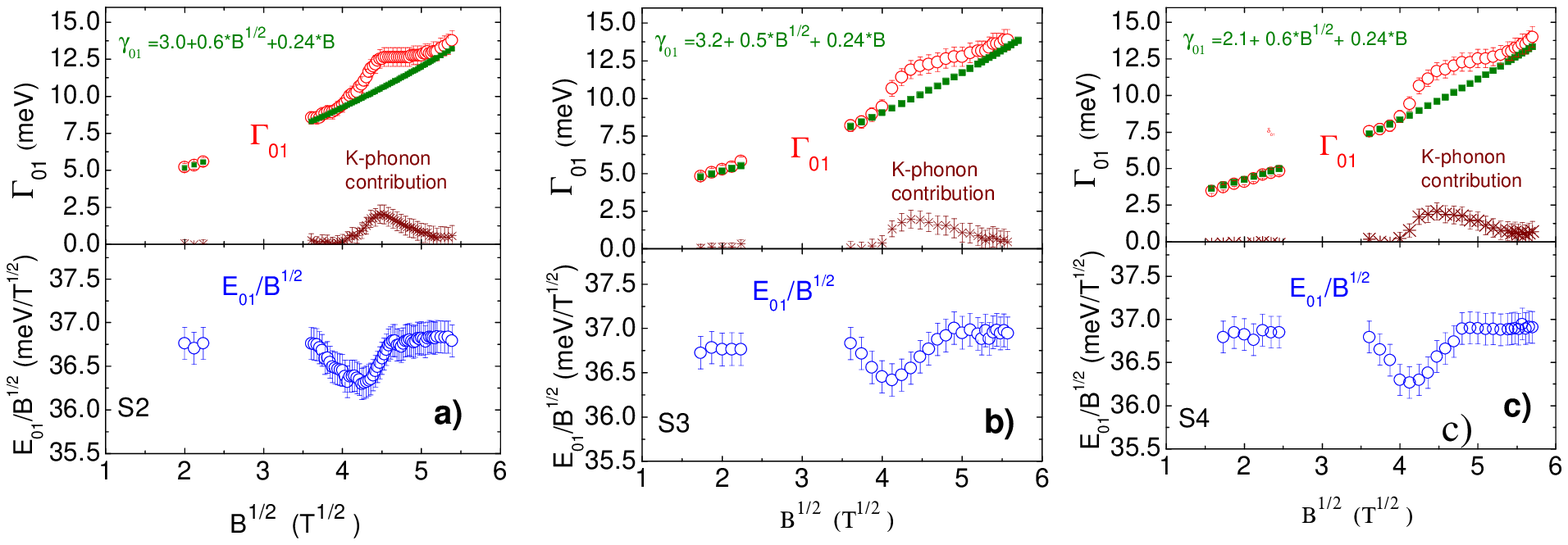}
\caption{\label{Fig2} Bottom panels: variation of
$E_{01}/\sqrt{B}$ as a function of $\sqrt{B}$ (open blue
circles).Top panels: variation of the fitted linewidth
$\Gamma_{10}(B)$ of the transition $E_{01}(B)$ as a function of
$\sqrt{B}$ (open red circles). This variation is decomposed is two
parts, one named $\gamma_{01}(B)$ (green dots) and the remaining
part ( wine stars) which will be assigned to the $K$-phonon
contribution. (a) sample S2, (b) sample S3  and (c) sample S4.}
 \end{figure*}

 One can, as said in the main paper, analyze globally the $E_{01}$
 transition by a fit with a single Lorentzian line which provides
 the evolution of
 the two parameters $E_{01}(B)$ and $\Gamma_{01}(B)$ as a function
 of $B^{1/2}$ as displayed in Fig.~\ref{Fig2} for samples S2, S3 and
 S4. However, as this will be explained in the following, the
 model for the electron-$K$-phonon interaction is dependent on the
 filling factor $\nu$ of the graphene layer and therefore on its carrier density
 $Ns$ ($\nu=Ns\Phi_{0}/B$, $\Phi_{0}$ being the flux quantum). We
 have therefore used  a multi-layer dielectric model to analyze the data. This model  assumes that each
graphene sheet is uniformly spread over the sample. This is a
strong assumption, difficult to justify \textit{a priori} but
which is necessary here to get some relative quantitative
information. If one knows, for a given sample, the number
$N_{eff}$ of effective layers and their carrier densities, one can
fit quantitatively the data using Eq.1 of the main text when
$v_{0}$, the Fermi velocity entering the optical matrix element
($M_{r,s}\propto v_{0}$), is known. Very often the value $v_{0}$
is taken identical to the Fermi velocity $v_{F}$ deduced from the
magnetic field dependence of the optical transitions. However
$v_{F}$ is the re-normalized velocity including electron-electron
interactions and it has been shown \cite{Bychkov} that $v_{0}$
should be the band velocity ignoring these interactions as given,
for instance, by LDA calculations. In the present work, we have
taken for all samples $v_{0}= 0.85\times 10^{6}m/sec$. We next
evaluate the number $N_{eff}$ for each sample. In the range of
magnetic fields 13 to 17 T, the relative transmission spectra
(Fig.~\ref{Fig1}) reaches values above 1 which depends on the
number $N_{eff}$: we have therefore a guide to estimate this
quantity. This is however only a guide because this value is also
sensitive to the effective coverage of each layer over the sample
and, of course, to experimental uncertainties. We estimate
$N_{eff}=3$ for sample S2, $N_{eff}=5$ for sample S3 and
$N_{eff}=6$ for samples S1 and S4. The carrier density $N_s$ for
each layer  is determined in the following way: one knows that,
for $2<\nu<6$, upon increasing $B$, the intensity of the $E_{01}$
absorption starts to increase, at the expense of the intensity of
the $E_{12}$ transition ($E_{12}=E_{2}-E_{1}$). The intensity does
not change with $B$ for $\nu<2$. Therefore, the disappearance of
the optical transition $E_{12}$ corresponds to $\nu=2$. Following
step by step, as a function of $B$, the transmission spectra one
can evaluate the carrier density $N_{sm}$ for each layer $m$. This
is a loop process which converges reasonably (within 20 per cent)
but has to be done independently for each sample. This gives
reliable values for the uppermost layers which are the less doped
and will play a major role when the electron-$K$-phonon
interaction will be switched on. The value of $N_{s1}$ for the
layer close to the SiC substrate can be set arbitrary to 5 to 6
$10^{12}cm^{-2}$ as given by transport data  on samples grown
under similar conditions: this layer indeed does not contribute to
the transition $E_{01}$ in the present experiment. Fitting the
data with such a multi-layer dielectric model and using Eq.1 of
the main text provides also the two parameters $E_{01}(B)$ and
$\Gamma_{01}(B)$: it turns out that within the experimental errors
(error bars in Fig.~\ref{Fig2}) they are identical to the results
obtained previously by fitting the data with a single Lorentzian
line. Therefore these results are quite robust.

We note that, for all samples, the variation of $\Gamma_{01}(B)$
as a function of $\sqrt{B}$ (top panels of Fig.~\ref{Fig2})
displays with respect to a smooth variation, an extra bump. We
have decomposed this variation into two parts, $\gamma_{01}(B)$
which is quadratic in $\sqrt{B}$ and an extra contribution
displayed by stars in the top panels of Fig.~\ref{Fig2}. We will
assign this latter contribution to the $K$-phonon interaction
contribution. We have taken for $\gamma_{10}(B)$ a quadratic
dependence of the form:
$\gamma_{01}(B)=\gamma_0+\gamma_1\times\sqrt{B}+\gamma_2\times B$.
Though it is possible to fit $\Gamma_{01}(B)$ for $B$ lower than
16 T with $\gamma_{2}=0$, the resulting decomposition gives, for
the $K$-phonon contribution, a variation which does not decreases
at high fields (or high energies) which is not physical. We are
then lead to introduce this $B$-dependence which, in turn, will
determine the way the interaction decreases at high energies. This
is of course dependent of the model used to describe the
interaction as discussed in section V-B. We have used the same
parameter $\gamma_{2}=0.24$ for all samples which in turn gives a
value for $\gamma_{1}=0.55 \pm0.05$ (the error bars covering the
coefficients for all samples).  This dependence on $B$ is  not
explained by existing scattering mechanisms \cite{Yang}. Coming
back to the $K$-phonon interaction contribution, we find, for all
samples a similar variation: after a threshold around 17 T, it
reaches a maximum of $2.1 \pm 0.1 meV$  (the error bars covering
the values found for all samples).

In the bottom panels of Fig.~\ref{Fig2} (and Fig 1b of the main
paper),  the variation of $E_{01}/B^{1/2}$ versus $B^{1/2}$ is
displayed. These variations show a downward kink at a value of $B$
corresponding to the onset of the $K$-phonon contribution. It is
important to note that, within the experimental errors, the
$K$-phonon contribution to $\Gamma_{01}(B)$ and the variation of
$E_{01}/B^{1/2}$, plotted as a function of the energy are indeed
Kramers-Kr\"{o}nig related.

Though the present analysis did not focus on the higher energy
transitions, we have fit as well over a large energy range the
transitions between the $n=-2$ to $n=1$ and $n=-1$ to $n=2$ Landau
levels corresponding to the energy
$\epsilon_{-1,2}=\epsilon_{-2,1}=v_{F}\sqrt{2e\hbar B}\times
(\sqrt{2}+1)$. For all samples this transition can be fitted with
$v_{F}=1.025\times 10^{6}\textrm{ms}^{-1}$. It is  interesting to
note that the width of  these transitions $\Gamma_{-12}(B)$ is
found to vary linearly as a function of $B^{1/2}$ like
$\Gamma_{-12}(B)= \eta_0+\eta_1\times \sqrt{B}$ without, within
the experimental errors, a significant contribution proportional
to $B$.  Therefore it seems that the dependence on $B$  adopted
for $\gamma_{01}(B)$ is specific of the $n=0$ LL.

\section{Fitting the experimental transmission curves}

Up to this point the treatment of the transmission curves has been
done assuming that all optically active layers have the same
broadening parameter $\Gamma_{10}(B)$. In reality the variation of
$E_{01}/B^{1/2}$ displayed in the bottom panels of Fig.~\ref{Fig2}
lead us to conclude that this $K$-phonon interaction, reminiscent
of the Fr\"{o}hlich interaction observed in polar semiconductors
\cite{Faugeras,Orlita}, has to be treated as a function of the
energy. In other words, for each layer simulated with Eq.1 of the
main text, the denominator $\hbar\omega -E_{01}(B)+ \imath
\Gamma_{01}(B)$  has to be replaced by $\hbar\omega
-\epsilon_{01}(B)+ \imath \gamma_{01}(B)+ \Sigma(\omega,B)$.
Therefore the new pole occurs at an energy
$\epsilon_{01}(B)-Re(\Sigma(\omega,B))$ with a width
$\gamma_{01}(B) +Im(\Sigma(\omega,B))$. Here $\Sigma(\omega,B))$
is the self energy of the phonon interaction and
$\epsilon_{01}(B)=v_{F}\sqrt{2e\hbar B}$ the non re-normalized
value of the transition energy. The model for treating the self
energy will be detailed in section V: the imaginary part of the
self energy depends on a parameter R (see Eq.\ref{ip}), the phonon
energy $\hbar\omega_{K}$ and an amplitude squared $A^{2}$
governing the electron-phonon interaction  (see section V-B). For
each sample we have used the background broadening parameter
$\gamma_{01}(B)$ as obtained from the preceding fitting (which is
sample dependent) but the parameters for the phonon-interaction
are the same for all samples: $R=3$, $\hbar\omega_{K}=151
~\textrm{meV}$ and $A^{2}=1.94~ \textrm{meV}^{2}\times B$. The
same Fermi velocity $v_{F}=1.012\times 10^{6}\textrm{ms}^{-1}$ has
been taken for all samples to describe the transition
$\epsilon_{01}(B)$. It is a little bit smaller than the value
found for the transition $\epsilon_{-12}(B)$ which may be a sign
of electron-electron interactions \cite{Bychkov}.

\begin{figure*}
\includegraphics*[width=0.9\columnwidth]{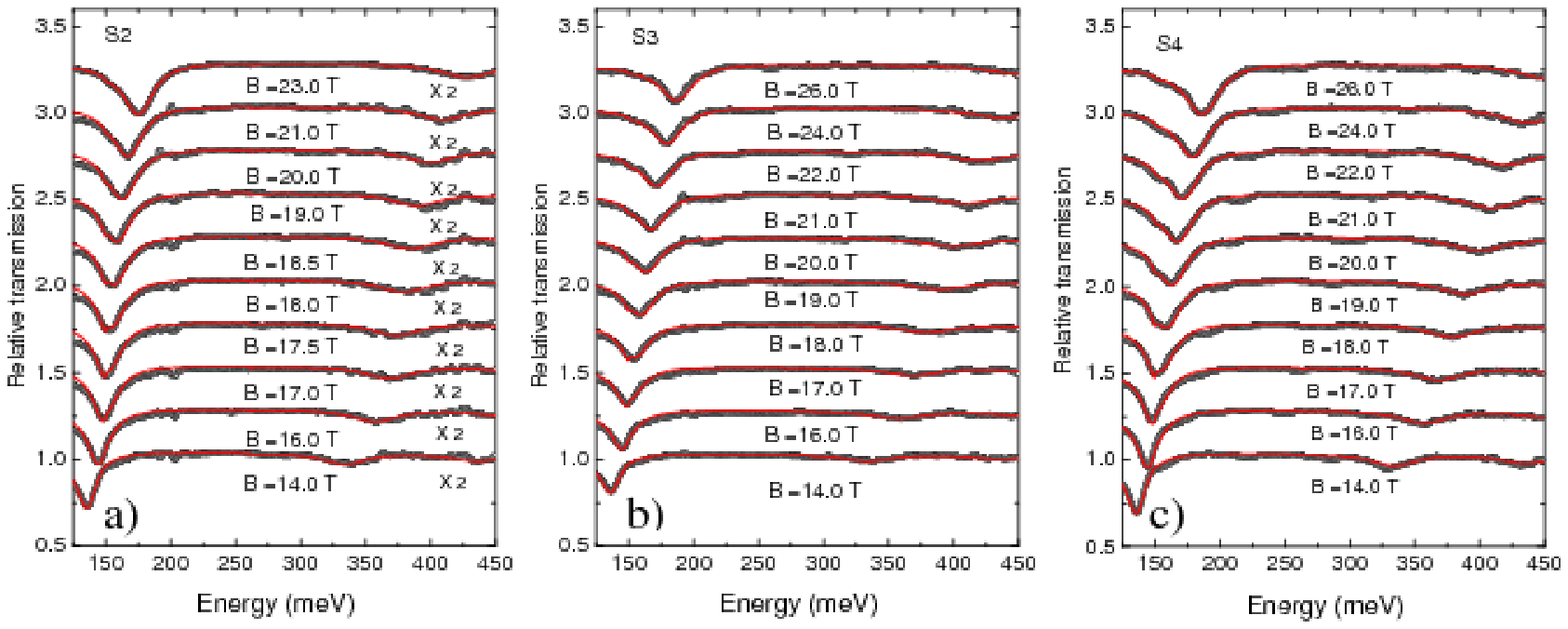}
\caption{\label{Fig3} Comparison of experimental data (open black
dots) with the simulated traces (red lines) using the proposed
K-phonon interaction model for different values of the magnetic
field. Theoretical spectra (from SI-IV) were calculated with parameters
described in the text. (a) sample S2 (scale multiplied by 2), (b)
sample S3 and (c) sample S4. }
 \end{figure*}

The results from the model, presented in the next sections, are
compared to the experimental data in Fig.~\ref{Fig3}. Including
the experimental errors, the parameter $R$ is determined within 15
per cent. However the fitting is done assuming that the valley
splitting $\Delta_{V}$ is larger than the spin splitting
$\Delta_{S}$, a result which is clearly beyond the experimental
uncertainties as evidenced by the comparison of both simulations
with the experimental spectra (see Fig.4 of the main text). As a
matter of fact, our model predicts, see section IV-B, that the
$K$-phonon interaction should vanish when the filling factor $\nu$
goes to zero if $\Delta_{V}<\Delta_{S}$.

Despite the different approximations made to treat the spectra, we
obtain a fit of transmission curves which is acceptable with  the
same basic parameters ($v_{0}$, $v_{F}$, R, $\hbar\omega_{K}$,
$A^{2}$) for all samples. This makes us confident in the basic
physical picture and the process we have adopted to treat these
spectra.

\section{Derivation of the model Hamiltonian}

\subsection{Electron-phonon interaction}

Expressions for the electron-$K$-phonon interaction in graphene in
the plane wave basis can be found in Ref.\cite{AndoKPhonon}. After
performing a change of basis to the Landau level (LL) basis, the
electron-phonon interaction can be written as:

\begin{equation}\label{eph}
H^{e-ph}=\sum_{\substack{j n n'\\
                         k k' \vec{q}}}\,
  g_{k k' \vec{q}}^{j n n'}\,c_{j+1,n',k'}^{\dagger}\,c_{j,n,k}\,
  (a^{\dagger}_{j,\vec{q}}+a_{j,\vec{q}}).
\end{equation}

Here, $c^{\dagger}_{j,n,k}$ is the electron creation operator in
valley $j=0,1$ (corresponding to the $K$ or $K'$ valley), for the LL $n$ and wavevector $k$
in the Landau gauge, and $a^{\dagger}_{j,\vec{q}}$ is the $K$ or
$K'$ phonon creation operator of wavevector $\vec{q}$ relative to
the $K$ or $K'$ point. The addition of the valley index $j+1$ is
understood to be modulo $2$. The matrix elements are:

\begin{equation}\label{matele}
g_{k k' \vec{q}}^{j n n'}=
  \frac{2\beta\gamma}{b^2}\,\sqrt{\frac{\hbar}{2N_cM\omega_K}}\,C
  \,i\,e^{2\pi{}i/3}\,\delta_{k,k'+q_x}\,e^{iq_y(k-q_x/2)l_B^2}\\
  \left(J_{|n'|-1,|n|-1}(\vec{q})-\textrm{sgn}(n)\textrm{sgn}(n')J_{|n'|,|n|}(\vec{q})\right)
\end{equation}

with

\begin{equation}
\begin{aligned}
J_{n',n}(\vec{q})&=
  \left(\frac{n'!}{n!}\right)^{1/2}\,e^{-q^2l_B^2/4}\,
  \left(\frac{q_xl_B+iq_yl_B}{2}\right)^{n'-n}L_n^{n'-n}\left(\frac{q^2l_B^2}{2}\right)
  \quad \textrm{when}\;n'\ge{}n\ge0\\
J_{n',-1}(\vec{q})&=0\\
J_{n',n}(\vec{q})&=J_{n,n'}^{*}(-\vec{q})\\
\end{aligned}
\end{equation}

In these equations, $b=a/\sqrt{3}$ is the C-C bond length,
$a=2.46~\AA$ is the lattice period, $\gamma=\sqrt{3}at_0/2$, and
$t_0=2.7 ~\textrm{eV}$ is the nearest neighbor hopping parameter.
$\beta=-(b/\gamma)(d\gamma/db)$ is a dimensionless coupling
parameter. $N_c$ is the number of unit cells, $M$ is the mass of a
carbon atom, $\omega_K$ is the $K$-phonon frequency and
$l_B=\sqrt{\hbar/(eB)}$ is the magnetic length. The constant $C$
takes the value of $1/2$ if $|n|>0$ and $|n'|>0$, and $1/\sqrt{2}$
otherwise. $L_m^n(u)$ denotes the Laguerre polynomial and
$\textrm{sgn}(n)=1$ when $n\ge1$ and  $\textrm{sgn}(n)=-1$ otherwise.

\subsection{Choice of reduced Hilbert space}

The cyclotron transitions of interest are (see Fig. 2 of main text):

\begin{equation}\label{eta}
\begin{aligned}
\lvert\eta_1(j,k)\rangle&=c^{\dagger}_{j,1,k}\,c_{j,0,k}\lvert0\rangle\\
\lvert\eta_2(j,k)\rangle&=c^{\dagger}_{j,0,k}\,c_{j,-1,k}\lvert0\rangle.
\end{aligned}
\end{equation}

Here, $\lvert0\rangle$ is the ground state. We now consider
candidate states $\lvert\xi\rangle$ to be included in the model,
such that the experimental spectra could be explained by their
interactions with the cyclotron transition states. Such states
have to satisfy two conditions: (i) they can couple directly to
$\lvert\eta_1\rangle$, $\lvert\eta_2\rangle$ via the
electron-phonon interaction Eq. \ref{eph} (i.e.
$\langle\xi\rvert{}H^{e-ph}\lvert\eta_{1,2}\rangle\ne0$), and (ii)
in order to describe the resonance at $E=\hbar\omega_K$ the states
$\lvert\xi\rangle$ must have energy
$\langle\xi\rvert{}H\lvert\xi\rangle\approx\hbar\omega_K$.

Based on the form of the electron-phonon interaction, it is
evident that $\lvert\xi\rangle$ must contain a $K$-phonon as well
as intervalley electronic transitions. Condition (ii) implies that
these transitions must be between LL of the same LL index because
the electronic part of $\lvert\xi\rangle$ must have the same
energy as $\lvert0\rangle$. We therefore consider the states

\begin{equation}\label{xi}
\lvert\xi_{\vec{q}}(j,k)\rangle=c^{\dagger}_{j+1,0,k-q_x}\,
  c_{j,0,k}\,a^{\dagger}_{j,\vec{q}}\lvert0\rangle.
\end{equation}

Note that the single phonon states $a^{\dagger}_{j,\vec{q}}\lvert0\rangle$
satisfy condition (ii) but not condition (i). In Eq. \ref{xi}, the
inter-valley transitions are between the $n=0$ LL because this is
the only partially empty LL in the experiment.

Another class of states that can couple to the cyclotron transition states
is\\*
$\{b^{\dagger}\lvert\xi_{\vec{q}}(j,k)\rangle,b'^{\dagger}b^{\dagger}\lvert\xi_{\vec{q}}(j,k)\rangle,\ldots\}$, which contain one
or more acoustic phonons in addition to
$\lvert\xi_{\vec{q}}(j,k)\rangle$.
Here, $b^{\dagger}$, $b'^{\dagger}$ create acoustic phonons.

\subsection{Integration over the phonon continuum}

Because of the dispersion of optical phonons at the $K$ point
\cite{Gruneis}, and the coupling to acoustic phonons, the energies
of the states that can couple to $\lvert\eta_1\rangle$,
$\lvert\eta_2\rangle$ form a continuum, bounded below by $
\omega_K$. In this section, we perform an integration over this
continuum.  We start with a Hamiltonian of a general form that
describes two degenerate states interacting with a continuum of
states. For a given $(j,k)$,

\begin{equation}\label{H1}
H_1=
\begin{pmatrix}
E_{01} & 0 & u_1( \omega_K, \omega_K) & u_1( \omega_K+\delta, \omega_K) & u_1( \omega_K+2\delta, \omega_K) & \dotso \\
0 & E_{01} & u_2( \omega_K, \omega_K) & u_2( \omega_K+\delta, \omega_K) & u_2( \omega_K+2\delta, \omega_K) & \dotso \\
u_1( \omega_K, \omega_K) & u_2( \omega_K, \omega_K) &  \omega_K & v & v & \dotso \\
u_1( \omega_K+\delta, \omega_K) & u_2( \omega_K+\delta, \omega_K) & v &  \omega_K+\delta & v & \dotso \\
u_1( \omega_K+2\delta, \omega_K) & u_2( \omega_K+2\delta, \omega_K) & v & v &  \omega_K+2\delta & \dotso \\
\vdots & \vdots & \vdots & \vdots & \vdots & \ddots
\end{pmatrix}.
\end{equation}

The first two columns (rows) represent the cyclotron transition
states $\lvert\eta_1\rangle$, $\lvert\eta_2\rangle$. The rest of
the columns (rows) represent the continuum of states that contain
phonons. Here, $\delta>0$ is a small positive energy. The function
$u_i( \omega, \omega_K)$ (may be obtained from Eqs. \ref{eph}, \ref{matele}) describes the interaction between
$\lvert\eta_i\rangle$ and a $\lvert\xi\rangle$ state of the continuum of energy
$ \omega$, while $v$ describes the interaction within the continuum.

We now treat the problem in two successive approximations. The
first one assumes that the phonons have no dispersion, which
allows us to introduce an effective coupling constant $V$ and work in a
restricted Hilbert space having effectively one phonon. The
second assumption takes into account the dispersion of phonons as
a correction to the first approximation. This is done by
introducing a self energy term in the Green's function (see next
section for the treatment of the Green's function).

An effective Hamiltonian in our restricted Hilbert space for
dispersion-less phonons can be written as:

\begin{equation}\label{H}
H=
\begin{pmatrix}
E_{01} & 0 & V_1 \\
0 & E_{01} & V_2 \\
V_1^* & V_2^* &  \omega_K
\end{pmatrix}
\end{equation}

\noindent
where the values of $V_1$ and $V_2$ are chosen such that $H$ and $H_1$
have the same eigenvalues when $\delta=0$. We have found that
$V_1=-V_2=V=\sqrt{\sum_{\vec{q}}{}'g_{1\vec{q}}\,g_{1\vec{q}}^*}$ (see
below for explicit expressions of this parameter.)

We then introduce the effects of the phonon continuum using second-order
L\"{o}wdin perturbation theory \cite{Lowdin}. This is done by
introducing a self energy $\Sigma_{ph}( \omega)$ in our Green's function
treatment of the optical response functions in the restricted Hilbert
space of $H$  \cite{Toyozawa}. The
condition imposed by the model is that the Green's functions
$(\omega-H-\Sigma_{ph}( \omega)-i0^+)^{-1}$ and
$(\omega-H_1-i0^+)^{-1}$ are equal in the space spanned by
$\lvert\eta_1\rangle$, $\lvert\eta_2\rangle$. Explicit expressions for
the different components of $\Sigma_{j}( \omega)$ are given in the
next section.

In the definition of the parameter $V$, the prime ($'$) on the sum
indicates that not all $\vec{q}$ vectors have to be summed over,
because only a fraction $0<f<1$ of the total number of
$\lvert\xi_{\vec{q}}\rangle$ states are available, depending on
the filling factor. Then,

\begin{equation}\label{V}
V=\sqrt{\sum_{\vec{q}}{}'g_{1\vec{q}}\,g_{1\vec{q}}^*}
  =\sqrt{f\sum_{\textrm{all }\vec{q}}g_{1\vec{q}}\,g_{1\vec{q}}^*}
  =A\sqrt{f}
\end{equation}

The constant $A^2=\sum_{\textrm{all
}\vec{q}}g_{1\vec{q}}\,g_{1\vec{q}}^*
  =\frac{9\sqrt{3}\beta^2\gamma^2B}
    {4\pi{}a^2M\omega_K}$ can be obtained from Eq. \ref{matele}. In reality this has
    to be corrected by the introduction of electron-electron interactions \cite{Gruneis}. To
obtain the value of $f$, we examine the occupation numbers of the
$n=0$ LL of the $K$ and $K'$ valleys.

\begin{figure}
\includegraphics*[width=0.8\columnwidth]{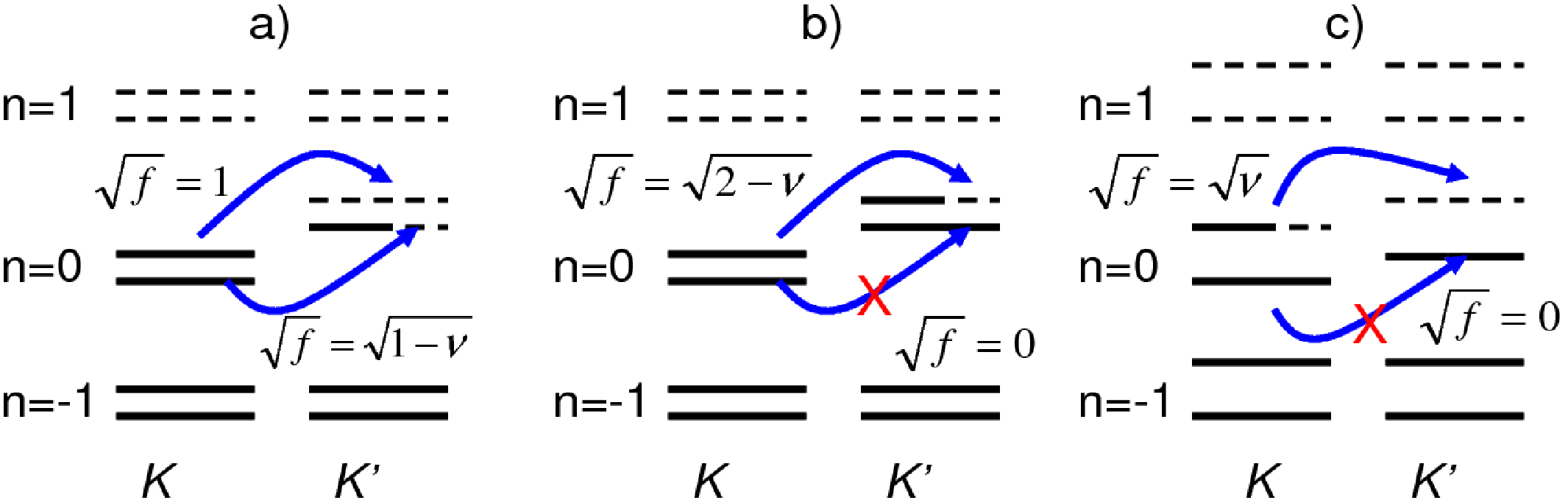}
\caption{\label{spval} a) Schematic diagram of the Landau levels
when $\Delta_V>\Delta_S$ and $\nu<1$; b) Schematic diagram of the
Landau levels when $\Delta_V>\Delta_S$ and $1<\nu<2$ and c)
Schematic diagram of the Landau levels when $\Delta_V<\Delta_S$
and $\nu<1$.}
\end{figure}

Let us define $\nu$ to be the filling factor, including spin and
orbital degeneracies. For example, $\nu=0$ at charge neutrality
and $\nu=2$ if the carrier concentration is such that the LL $n=0$
is fully occupied. For the case of $\Delta_V>\Delta_S$, filling
factor $\nu<1$ (Fig. \ref{spval}a), we have $\sqrt{f}=1$ for the
spin up transitions, and $\sqrt{f}=\sqrt{1-\nu}$ for the spin down
transitions. Whereas for $\Delta_V>\Delta_S$, filling factor
$1<\nu<2$ (Fig. \ref{spval}b), we have $\sqrt{f}=\sqrt{2-\nu}$ for
the spin up transitions, and $\sqrt{f}=0$ for the spin down
transitions.

On the other hand, if $\Delta_V<\Delta_S$, the expressions are the
same as the case of $\Delta_V<\Delta_S$ when the filling factor is
$1<\nu<2$, but when $\nu<1$: we have $\sqrt{f}=\sqrt{\nu}$ for the
spin up transitions, and $\sqrt{f}=0$ for the spin down
transitions (Fig. \ref{spval}c). Therefore, in that case, for
undoped graphene layer, the interaction disappears.

\section{Evaluation of optical conductivity}

\subsection{Green function formalism}

The Green function formalism for the optical conductivity
$\textrm{Re}\,\sigma_{ij}\sim\textrm{Im}\,\langle{}0\lvert{}M_{i}^\dagger{}GM_{j}\rvert{}0\rangle$
has been
introduced in the main text. Here, we provide some explicit expressions
for the form of the optical conductivity. In order to facilitate the
discussion of the self energy $\Sigma_{ph}( \omega)$, we express the Green's
function matrix in the basis of eigenstates of the model Hamiltonian
$H$ (Eq. \ref{H}).

\begin{equation}\label{diagG}
G'=
\begin{pmatrix}
G_{11}' & 0 & 0\\
0 & G_{22}' & 0\\
0 & 0 & G_{33}'
\end{pmatrix},\quad
G_{jj}'=\frac{1}{\omega-\varepsilon_j-is},\quad{}s=0^+
\end{equation}

The primed (') quantities are expressed in the basis of the eigenstates
of $H$. The
eigenvalues $\varepsilon_j$ of the Hamiltonian $H$ (Eq. \ref{H}) are

\begin{equation}\label{eigenvalues}
\begin{aligned}
\varepsilon_1&=E_{01}\\
\varepsilon_{2,3}&=\frac{1}{2}\left(E_{01}+\hbar\omega_K\pm\sqrt{(E_{01}-\hbar\omega_K)^2+8V^2}\right)
\end{aligned}
\end{equation}.

\noindent
The Green's function matrix ($G$) in the original basis (i.e., the same basis
as Eq. \ref{H}) can be recovered by a unitary transformation
$G=UG'U^{-1}$ .

\begin{equation}\label{U}
U=
\begin{pmatrix}
\frac{1}{\sqrt{2}}&\frac{-V}{\sqrt{2V^2+(E_{01}-\varepsilon_2)^2}}&\frac{-V}{\sqrt{2V^2+(E_{01}-\varepsilon_3)^2}}\\
\frac{1}{\sqrt{2}}&\frac{V}{\sqrt{2V^2+(E_{01}-\varepsilon_2)^2}}&\frac{V}{\sqrt{2V^2+(E_{01}-\varepsilon_3)^2}}\\
0&\frac{E_{01}-\varepsilon_2}{\sqrt{2V^2+(E_{01}-\varepsilon_2)^2}}&\frac{E_{01}-\varepsilon_3}{\sqrt{2V^2+(E_{01}-\varepsilon_3)^2}}
\end{pmatrix}
\end{equation}

\noindent

 Then, for instance, the $\sigma_{xx}$ component of the optical
conductivity is related to
$\langle{}0\lvert{}M_{x}^\dagger{}GM_{x}\rvert{}0\rangle
=\langle{}0\lvert{}M_{x}^\dagger{}UG'U^{-1}M_x\rvert{}0\rangle$,
where $M_x^\dagger=(M_{1x},M_{2x},0)$ is a row vector containing
the optical matrix elements.

\subsection{Discussion of $\Sigma_{ph}$}

We introduce self energy terms into the diagonal elements of $G'$

\begin{equation}\label{replace}
G_{jj}'=\frac{1}{\omega-\varepsilon_j-i0^+}\rightarrow\frac{1}{\omega-\varepsilon_j-\Sigma_j(\omega)}
\end{equation}

From second-order L\"{o}wdin perturbation theory (see previous section),
we have found
\begin{equation}\label{sigma}
\begin{aligned}
\textrm{Im}\,\Sigma_1&=0\\
\textrm{Im}\,\Sigma_2&=\Sigma(\omega,\omega_K)\\
\textrm{Im}\,\Sigma_3&=\Sigma(\omega,\omega_K)
\end{aligned}
\end{equation}

\noindent
where the imaginary part of the self energy has the form

\begin{equation}\label{imsigma}
\textrm{Im}\,\Sigma( \omega)=\theta( \omega- \omega_K)(u( \omega, \omega_K))^2
\end{equation}

\noindent
Here, $u(\omega,\omega_K)=\lvert{}u_1(\omega,\omega_K)\rvert=\lvert{}u_2(\omega,\omega_K)\rvert$.
In the basis of Eq. \ref{H}, the Green's function  can
be written as

\begin{equation}\label{G}
G=
\begin{pmatrix}
G_{11} & G_{12} & G_{13}\\
G_{21} & G_{22} & G_{23}\\
G_{31} & G_{32} & G_{33}
\end{pmatrix}
\end{equation}

\begin{figure}
\includegraphics*[width=0.5\columnwidth]{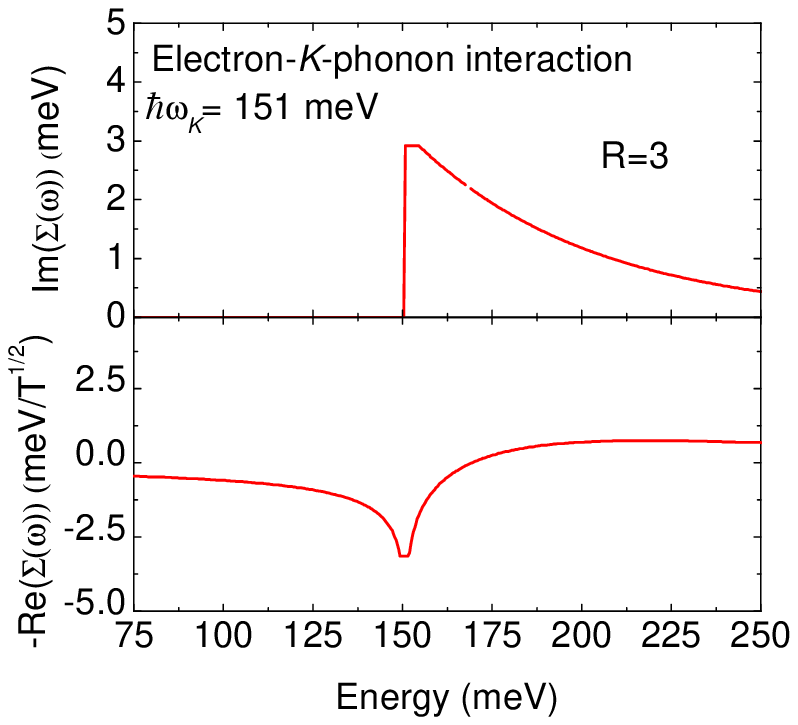}
\caption{\label{KK} Top panel: Imaginary part of the self-energy
($\Sigma_2$ or $\Sigma_3$ in Eq. \ref{sigma}) with $\nu=0$. Bottom
panel: Real part of the self-energy with $\nu=0$ as deduced from
Kramers-Kronig transformation of the Imaginary part. The behavior
of the real part of the self-energy, as a function of energy, is
quite similar to that of $E_{01}/B^{1/2}$ when adding
 the contribution of the  non re-normalized
value of the transition energy
$\epsilon_{01}(B)/\sqrt{B}=v_{F}\sqrt{2e\hbar}= 36.8 meV/T^{1/2}$.
(See bottom panels of Fig.~\ref{Fig2} and Fig.1b of the main
text.)}
\end{figure}

\begin{equation}\label{Gelements}
\begin{aligned}
G_{11}&=G_{22}=\frac{1}{2}\left(\frac{1}{ \omega-E_{01}-i0^+}+\frac{1}{ \omega-E_{01}-i0^+-\Sigma( \omega)}\right)\\
G_{12}&=-\frac{1}{2}\left(-\frac{1}{ \omega-E_{01}-i0^+}+\frac{1}{ \omega-E_{01}-i0^+-\Sigma( \omega)}\right).
\end{aligned}
\end{equation}

$\textrm{Im}\,\Sigma( \omega)$ has a threshold at $ \omega_K$ that
arises irrespective of the form of the coupling matrix element
$u$. In this model, the threshold is always located at the lower
bound of the energies of the continuum of states. We have chosen,
motivated by the experimental data (see top panels of
Fig.~\ref{Fig2}), for $u^2$ the form

\begin{equation}\label{ip}
(u( \omega, \omega_K))^2=D\,\textrm{exp}(-R(\omega/\omega_K-1))
\end{equation}

\noindent

Using the sum rule for $\textrm{Im}(G)$ together with the
constraint that $H_1$ must reduce to $H$ in the limit where
$R\rightarrow\infty$ (the limit where the continuum collapses to a
single energy $ \omega_K$), the value of $D$ can be fixed at
$D=(\pi{}R\,V^2)/(\hbar\omega_K)$. The model is not sensitive to
the exact functional form of Eq. \ref{imsigma} as long as the
qualitative features such as the exponential decay and the
presence of a threshold at $\omega=\omega_K$ are preserved. We
then have two independent parameters to describe the interaction,
$V^{2}$ and $R$. As derived in Eq. \ref{V}, $V^{2}=A^{2}f$ and we
will use the value $A^{2}(~\textrm{meV}^2)=1.94 \times
B(\textrm{T})$ \cite{Gruneis}. The energy range of
$u(\omega,\omega_K)$ contains contribution from the $K$-phonon
dispersion and multi-acoustic phonon processes. The only remaining
fitting parameter is $R$. In the interpretation of the continuum
as states containing multiple acoustic phonons, the exponential
factor in Eq. \ref{ip} describes the decreasing probability of
multiple acoustic phonon emission far from the threshold energy.
With this simplified model, and knowing the experimental results,
we have chosen the value $R=3$ as sketched in Fig. \ref{KK} which
in turn is coherent with the value of the linear $B$ term used for
$\gamma_{01}(B)$. This also gives reasonable values for
$\textrm{Re}\,\Sigma( \omega)$ (bottom panel Fig. \ref{KK}) as
compared to the experimental values. Of course the extension in
energy corresponding to this $R$ value is quite high but it does
not play a role in the actual fitting process limited to about 200
meV (see  Fig. \ref{Fig3}). Though, within the experimental
errors, the fit of data is acceptable with $R=3$, it could  be
also accepted for $R$ values, as well as the $\gamma_{2}$ values
of $\gamma_{01}(B)$  differing by 15 per cent. Knowing that the
spin splitting $\Delta_{S}/B\simeq 0.12-0.15$ meV/T and that the
the introduction of $\Delta_{V}\propto B$ should result in an
extra broadening of the $E_{01}$ transition, the introduction of
the $\gamma_{2}$-term is explained naturally by the fact that it
should  be larger than $\Delta_{S}/B$.

We note finally that the introduction of the interference term
$G_{12}$ (Eq.\ref{Gelements}) (which corresponds to interference
effects between the cyclotron transitions at the two valleys) is
essential to reproduce the experimental results.

\bibliography{basename of .bib file}